# A Novel Distance Matric: Generalized Relative Entropy

Shuai Liu[a,b,*], Mengye Lu[a,b], Gaocheng Liu[a,b], Zheng Pan[a,b]

a. College of Computer Science, Inner Mongolia University, Hohhot, China

b. Inner Mongolia Key Laboratory of Data Mining and Knowledge Engineering, Hohhot, China

* Corresponding author: cs_liushuai@imu.edu.cn

**Abstract.** Information entropy and its extension, which are important generalization of entropy, have been applied in many research domains today. In this paper, a novel generalized relative entropy is constructed to avoid some defects of traditional relative entropy. We presented the structure of generalized relative entropy after the discussion of defects in relative entropy. Moreover, some properties of the provided generalized relative entropy is presented and proved. The provided generalized relative entropy is proved to have a finite range and is a finite distance metric.

## 1. Background

The concept of entropy was proposed by T. Clausius as one of the parameters to reflect the degree of chaos for the object. Later, research found that information was such an abstract concept that is hard to make it clear to obtain its amount. Indeed, it was not until the information entropy was proposed by Shannon that we had a standard measure for the amount of information. Then, some related concepts based on information entropy has been proposed subsequently, such as cross entropy, relative entropy and mutual information, which offered an effective method to solve the complex problems of information processing. Therefore, the study of novel metric based on information entropy was significant in the research domain of information science.

Information entropy was first proposed by Shannon. Assuming an information source I have composed by n different signals $I_i$, H(I), the information entropy of I, was shown in Eq.1, where $p_i = \frac{amount\ of\ I_i}{signal's\ amount\ of\ I}$ denotes frequency of $I_i$, E() means mathematical expectation, $k > 1$ denotes the base of logarithm. When $k = 2$, the unit of H(I) is bit.

$$H(I) = E(-\log_k p_i) = -\sum_{i=1}^{n} p_i \cdot \log_k p_i \qquad (1)$$

Information entropy was a metric of the chao's degree for an information source. An information source was more complex when its information entropy was larger, and vice versa. The information entropy is an effective internal metric to measure an information source. To extend information entropy to measure the best distribution of an information source, cross entropy was defined in Eq.2 where P was 'real' information source and Q was a 'unreal' distribution information source, $p_i$ denotes frequency of components of P and $q_i$ denotes frequency of components of Q.

$$H(P,Q) = \sum_{i=1}^{n} p_i \cdot \log_k \frac{1}{q_i} \qquad (2)$$

Cross entropy also can act as the reaction of the similarity degree of component's distribution for the two information sources. $H(P,Q) = H(P)$ if and only if all components' distribution were identical. An homologous metric was relative entropy, and was also known as Kullback–Leibler divergence. Its definition was shown in Eq.3, where Eq 3.1 was relative entropy of the discrete random variables and Eq.3.2 was of the continuous random variables.

$$D(P||Q) = \sum_{i=1}^{n} p_i \cdot \log_k \frac{p_i}{q_i} \qquad (3.1)$$

$$D(P||Q) = \oint p_x \cdot \log_k \frac{p_x}{q_x} dx \qquad (3.2)$$

Relative entropy reflected the difference of the two information sources with different distributions, the larger relative entropy was, the difference of the two information sources was larger, vice versa. However, mutual information, another entropy based metric was also proposed to indicate the relationship of information which contained in two random variables X and Y. Mutual information was defined as relative entropy of p(x,y) and p(x)·p(y) in Eq.4, where f(x,y) was joint probability density, $f_X(x)$ was marginal probability density of X and $f_Y(y)$ was marginal probability density of Y. The value of mutual information was non-negative. If and only if X and Y were independent variables with each other, the value of mutual information was equal to zero.

$$I(X;Y) = \oiint_{x \in X, y \in Y} p(x,y) \cdot \log_k \frac{p(x,y)}{p(x) \cdot p(y)} \qquad (4)$$

Recently, there are many extensions and applications of information entropy based metric. But all these information entropy based methods have some defects. The most two important defects are: (1) it is not a distance metric; (2) it does not have an upper bound.

So in this paper, Section 2 introduces related works with information entropy based methods; Section 3 provides a novel generalized relative entropy and prove some properties of the provided entropy; Section 4 summarizes the whole content.

**2. Related work**

For years, many scholars have studied the applications of various entropy. Earlier, Iwo Białynicki-Birula et al. deduced new uncertain relationship in quantum mechanics based on information entropy [1]. A.Uhlmann et al. applied relative entropy in digital integration, and proved some property of the interpolation theory [2]. J. Shore et al. deduced the principle of maximum entropy and minimum cross entropy [3]. Andrew M. Fraser et al. analyzed the coordinate of singular factors [4]. Steven M.pincus et al. analyzed the complex degree of the system by the entropy [5]. Afterwards, Aapo Hyvarinen et al. analyzed independent component and projection pursuit based on entropy [6].

In 2000, I.R.Petersen et al. analyzed the optimization problem for the system with constraint of the relative entropy [7]. N.kwak et al. classified sample based on mutual information between the input information and the variable category [8]. Later, J.P.W.Pluim et al. analyzed the image matching in medicine based on mutual information [9]. M. Arif et al. used the entropy to analyze the changes of the center of gravity between the old and young in order to find a method of improving the walking stability for the old [10]. Steven J. Phillips et al. analyzed the distribution of species by maximum entropy model [11]. V Krishnaveni et al. analyzed the electroencephalogram of human by mutual information [12]. Afterwards, Wolf M M et al. researched area laws in quantum systems by used mutual information and correlations [13]. Roger A. Baldwin et al. utilized maximum entropy model to find some regular about the selection of habitat of wild animal [14]. Sergio Verdu et al.

combined the matching with relative entropy and analyzed the relationship both of them [15].

In 2011, Batina L et al. reviewed mutual information [16]. Audenaert K M studied the asymmetry of the relative entropy [17]. Maoguo Gong et al. made the best of the BIFT and mutual information to propose a method that can match the object precisely [18]. I. Giagkiozis et al. proposed a new method that can took advantage of the knowledge of cross entropy to solve the problem of multi-object programming [19]. Tang M and Mao X researched information entropy-based metrics for measuring emergences in artificial societies [20].

In recent years, a great number of scholars have studied entropy based methods in recognition and classification. Soares C and Knobbe A. studied entropy-based discretization methods for ranking data [21]. Zhengfu Li et al. proposed a method to solve the problem of molecular docking used information entropy and ant colony genetic algorithm [22]. Chun-Wang Ma et al. used information entropy to analyze the changes of substance in the process of chemical change [23]. Kö et al. researched operational meaning of min- and max-entropy [24]. Besides, Müller M P and Pastena M. studied a generalization of majorization based on Shannon entropy [25]. Xiao Zhang et al. proposed a feature selection algorithm for fuzzy rough sets on the basis of information entropy [26]. Abolfazl Ebrahimzadeh et al. proposed the concept of logical entropy based on entropy, and applied it to quantum dynamical system [27]. Pedro Lopez-Garcia et al. proposed a method to make a prediction for traffic jam in a short period of time combined the genetic algorithm with cross entropy [28]. David Sutter et al. studied the monotonicity of cross entropy [29]. Opper M provided an estimator for the relative entropy rate of path measures for stochastic differential equations. Tang L et al. studied an EEMD-based multi-scale fuzzy entropy approach for complexity analysis in clean energy markets.

## 3. Generalized Relative Entropy
### 3.1 Structure of Generalized Relative Entropy

Nowadays, the relative entropy becomes one of the most important dissimilarity measure between two multidimensional vectors. Let $X(x_1, \ldots x_s)$ and $Y(y_1, \ldots y_s)$ be two multidimensional vectors, which are constituted by s components with different counts. The count of $i^{th}$ component is $x_i$ in vector X and $y_i$ in vector Y. Therefore, the relative entropy RE(X, Y), which denotes relation from X to Y, is defined in Eq.5, where $p_x(i) \stackrel{\text{def}}{=} \frac{x_i}{\sum_{i=1}^{s} x_i}$ means the probability of $x_i$ in X for each i. Herein, we define $p_x(i) \cdot \log \frac{p_x(i)}{p_y(i)} = 0$ when $p_x(i) = 0$ in order to avoid form of equation $0 \cdot \log 0$. In real application, $\varepsilon > 0$ is added in the denominator of log(), which is a very small positive number to avoid form of equation $\log \infty$.

$$RE(X, Y) = \sum_{i=1}^{s} p_x(i) \cdot \log \frac{p_x(i)}{p_y(i)} \quad (1)$$

It is admittedly that RE(X, Y) is not a distance metric because usually $RE(X, Y) \neq RE(Y, X)$ when $X \neq Y$. However, relative entropy does not have finite upper bound, which makes it can not be easily used to measure difference between high dimensional vectors in real application. So in this paper, based on definition of relative entropy, we present a generalized relative entropy d(X, Y) by Eq.6, where s denotes the number of components and $k \geq 1$ denotes the control parameter of function d(), $r = 0$ when $X = Y$; $r = 1$ when $X \neq Y$. We believe the generalized relative entropy has better properties than relative entropy. Moreover, it is a distance metric.

$$d(X, Y) = \sum_{i=1}^{s} \left( p_x(i) \cdot \log \frac{k \cdot p_x(i)}{(k-1)p_x(i) + p_y(i)} + p_y(i) \cdot \log \frac{k \cdot p_y(i)}{p_x(i) + (k-1)p_y(i)} \right) + r$$
$$\cdot \log \left(1 + \frac{1}{k-1}\right)^2 \quad (6)$$

### 3.2 Properties of Generalized Relative Entropy

Theorem 1 will be presented to prove that the generalized relative entropy d() is a distance metric. However, Lemmas 1-2 and Inferences 1-2 are presented at first.

**Lemma 1.** $\sum_{i=1}^{s}\left(p_x(i) \cdot \log \frac{k \cdot p_x(i)}{(k-1)p_x(i)+p_y(i)}\right)$ is constant nonnegative if $p_x(i) \geq 0, p_y(i) \geq 0, \sum_{i=1}^{s} p_x(i) = 1, k \geq 1$.

**Proof.**

Because $p_x(i) \cdot \log \frac{k \cdot p_x(i)}{(k-1)p_x(i)+p_y(i)} = -p_x(i) \cdot \log \frac{(k-1)p_x(i)+p_y(i)}{k \cdot p_x(i)}$, $\frac{(k-1)p_x(i)+p_y(i)}{k \cdot p_x(i)} \geq 0$, we have following Eq.7.

$$\sum_{i=1}^{s}\left(p_x(i) \cdot \log \frac{(k-1) \cdot p_x(i) + p_y(i)}{k \cdot p_x(i)}\right) \leq \log \sum_{i=1}^{s} \frac{(k-1) \cdot p_x(i) + p_y(i)}{k \cdot p_x(i)} \cdot p_x(i)$$

$$= \log \frac{(k-1) \cdot \sum_{i=1}^{s} p_x(i) + \sum_{i=1}^{s} p_y(i)}{k} = \log 1 = 0 \quad (7)$$

So $\sum_{i=1}^{s} p_x(i) \cdot \log \frac{k \cdot p_x(i)}{(k-1)p_x(i)+p_y(i)} = -\sum_{i=1}^{s} p_x(i) \cdot \log \frac{(k-1)p_x(i)+p_y(i)}{k \cdot p_x(i)} \geq 0$

Lemma 1 is proved. ∎

Then, with consideration of condition that sign "=" appeared, we have Reference 1.

**Inference 1.** $\sum_{i=1}^{s}\left(p_x(i) \cdot \log \frac{k \cdot p_x(i)}{(k-1)p_x(i)+p_y(i)}\right)$ is zero if and only if $p_x(i) = p_y(i)$ for all i where $p_x(i) \geq 0, p_y(i) \geq 0, \sum_{i=1}^{s} p_x(i) = \sum_{i=1}^{s} p_y(i) = 1$ and $k \geq 1$.

Then, we have Lemma 2 and Inference 2 based on Lemma 1 and Inference 1 to prove upper bound of d(X,Y).

**Lemma 2.** $\sum_{i=1}^{s}\left(p_x(i) \cdot \log \frac{k \cdot p_x(i)}{(k-1)p_x(i)+p_y(i)}\right) \leq \log \frac{k}{k-1}$ if $p_x(i) \geq 0, p_y(i) \geq 0, k \geq 1, \sum_{i=1}^{s} p_x(i) = 1$.

**Proof.**

We have Eq.8 to prove Lemma 2 based on Eq.9.

$$\sum_{i=1}^{s}\left(p_x(i) \cdot \log \frac{k \cdot p_x(i)}{(k-1)p_x(i)+p_y(i)}\right) \leq \sum_{i=1}^{s}\left(p_x(i) \cdot \log \frac{k}{k-1}\right) = \log \frac{k}{k-1} \quad (8)$$

$$\log \frac{k \cdot p_x(i)}{(k-1)p_x(i)+p_y(i)} \leq \log \frac{k \cdot p_x(i)}{(k-1)p_x(i)} = \log \frac{k}{k-1} \quad (9)$$

Lemma 2 is proved. ∎

**Inference 2.** Upper bound of $d(X, Y)$ is $4 \cdot \log \frac{k}{k-1}$ where $\sum_{i=1}^{s} p_x(i) = \sum_{i=1}^{s} p_y(i) = 1$ and $k \geq 1$.

**Proof.**

We have Eq.10 to prove Inference 2 based on Lemma 2.

$$d(X, Y) = \sum_{i=1}^{s}\left(p_x(i) \cdot \log \frac{k \cdot p_x(i)}{(k-1)p_x(i)+p_y(i)} + p_y(i) \cdot \log \frac{k \cdot p_y(i)}{p_x(i)+(k-1)p_y(i)}\right) + r \cdot \log\left(1 + \frac{1}{k-1}\right)^2$$

$$\leq \log \frac{k}{k-1} + \log \frac{k}{k-1} + 2r \cdot \log \frac{k}{k-1} \quad (10)$$

Inference 2 is proved. ∎

After that, Theorem 1 is presented to prove d(X,Y) is a distance metric between two elements X and Y with same diversity s and length n.

**Theorem 1. Function d() is a distance metric of elemental set E{} in space S(E{}, d()) where all elements in E have same diversity s.**

**Proof.**

Let X and Y are two elements in E, $p_x(i)$ and $p_y(i)$ denotes frequency of $i^{th}$ component in X or Y, $k > 1$ is a control parameter, s is the number of components in X and Y, $r = 0$ when $X = Y$ and $r = 1$ when $X \neq Y$, we have $d(X,Y) = \sum_{i=1}^{s} \left( p_x(i) \cdot \log \frac{k \cdot p_x(i)}{(k-1)p_x(i)+p_y(i)} + p_y(i) \cdot \log \frac{k \cdot p_y(i)}{p_x(i)+(k-1)p_y(i)} \right) + r \cdot \log \left(1 + \frac{1}{k-1}\right)^2$ from Eq.2.

Then, we use following properties to prove Theorem 1.

**Property 1. $d(X,Y) \geq 0$ for every X and Y, $d(X,Y) = 0$ if and only if $X = Y$.**

First, we know $\sum_{i=1}^{s} \left( p_x(i) \cdot \log \frac{k \cdot p_x(i)}{(k-1)p_x(i)+p_y(i)} \right)$ is nonnegative from Lemma 1. It implies $\sum_{i=1}^{s} \left( p_y(i) \cdot \log \frac{k \cdot p_y(i)}{(k-1)p_y(i)+p_x(i)} \right) \geq 0$. Then, we know $r \cdot \log \left(1 + \frac{1}{k-1}\right)^2 \geq 0$. So $d(X,Y) = \sum_{i=1}^{s} \left( p_x(i) \cdot \log \frac{k \cdot p_x(i)}{(k-1)p_x(i)+p_y(i)} \right) + \sum_{i=1}^{s} \left( p_y(i) \cdot \log \frac{k \cdot p_y(i)}{p_x(i)+(k-1)p_y(i)} \right) + r \cdot \log \left(1 + \frac{1}{k-1}\right)^2 \geq 0$, which means Property 1 is proved.

**Property 2. $d(X,Y) = d(Y,X)$ for every X and Y.**

It is admittedly that the formation of $d(X,Y)$ is symmetrical to $d(Y,X)$ for every pair of vectors X and Y, which means Property 2 is proved.

**Property 3. $d(X,Y) + d(Y,Z) \geq d(X,Z)$.**

First, if there are at least two elements in $\{X,Y,Z\}$ are equal, it is admittedly that $d(X,Y) + d(Y,Z) \geq d(X,Z)$ because in the three functions d(), one value is zero and other two values are same and nonnegative.

So, Eq.11 is used to describe $d(X,Y) + d(Y,Z) - d(X,Z)$ when $X \neq Y \neq Z$.

$d(X,Y) + d(Y,Z) - d(X,Z)$

$$= \sum_{i=1}^{s} \left( p_x(i) \cdot \log \frac{k \cdot p_x(i)}{(k-1)p_x(i) + p_y(i)} + p_y(i) \cdot \log \frac{k \cdot p_y(i)}{p_x(i) + (k-1)p_y(i)} + p_y(i) \right.$$

$$\left. \cdot \log \frac{k \cdot p_y(i)}{(k-1)p_y(i) + p_z(i)} + p_z(i) \cdot \log \frac{k \cdot p_z(i)}{p_z(i) + (k-1)p_y(i)} - p_x(i) \right.$$

$$\left. \cdot \log \frac{k \cdot p_x(i)}{(k-1)p_x(i) + p_z(i)} - p_z(i) \cdot \log \frac{k \cdot p_z(i)}{p_x(i) + (k-1)p_z(i)} \right) + \log \left(1 + \frac{1}{k-1}\right)^2 \quad (11)$$

Then, we have Eq.12 to prove Property 3 based on Lemmas 1-2.

$d(X,Y) + d(Y,Z) - d(X,Z)$

$$\geq \log \left(1 + \frac{1}{k-1}\right)^2 - \sum_{i=1}^{s} p_x(i) \cdot \log \frac{k \cdot p_x(i)}{(k-1)p_x(i) + p_z(i)}$$

$$- \sum_{i=1}^{s} p_z(i) \cdot \log \frac{k \cdot p_z(i)}{p_x(i) + (k-1)p_z(i)} \geq 2 \cdot \log \frac{k}{k-1} - 2 \cdot \log \frac{k}{k-1} \geq 0 \quad (12)$$

To summarize Properties 1-3, Theorem 1 is proved.∎

Then, we have Theorem 2 to provide the range of d(X, Y) for all elements X and Y which combined with s components.

**Theorem 2.** Range of $d(X, Y)$ is $\{0\} \cup [2 \cdot \log\frac{k}{k-1}, 4 \cdot \log\frac{k}{k-1}]$.

**Proof.**

Case 1) $X = Y$

When $X = Y$, we have $d(X, Y) = d(X, X) = 0$ by Inference 1.

Case 2) $X \neq Y$

When $X \neq Y$, we have $d(X, Y) \leq 4 \cdot \log\frac{k}{k-1}$ by Inference 2, and $d(X, Y) \geq 2 \cdot \log\frac{k}{k-1}$ by Lemma 1.

To summarize Cases 1-2, Theorem 2 is proved.∎

In this way, the generalized relative entropy is provided with its some properties are proved.

**4. Conclusion**

In this paper, we provided a novel distance metric based on relative entropy, which was called generalized relative entropy. The generalized relative entropy surmounted the disadvantage of relative entropy that it had an upper bound and satisfies triangle inequality of distance. The properties of distance metric and upper bound were proved in this paper. Finally, range of the provided generalized relative entropy was computed. Since there was a parameter k to control the generalized relative entropy, we believed that this metric can be used in variety of real applications by adjusted k.